\documentstyle[12pt]{article}

\newcommand{\Diff}[2]{\frac{\partial #1}{\partial #2}}
\newcommand{\CD}[1]{\delta_{#1}}
\newcommand{\PB}[1]{\left\{#1\right\}_{\rm P}}
\newcommand{\DB}[1]{\left\{#1\right\}_{\rm D}}
\newcommand{\XX}{x_1^2+x_2^2}
\newcommand{\OXX}{\hat{x}_1^2+\hat{x}_2^2}
\newcommand{\RXX}{\sqrt{\XX}}
\newcommand{\ORXX}{\sqrt{\OXX}}
\newcommand{\XP}{x_1P_1+x_2P_2}
\newcommand{\OXP}{\hat{x}_1\hat{P}_1+\hat{x}_2\hat{P}_2}
\newcommand{\XD}[1]{x_1\CD{1#1}+x_2\CD{2#1}}
\newcommand{\OXD}[1]{\hat{x}_1\CD{1#1}+\hat{x}_2\CD{2#1}}
\newcommand{\RpaS}{R+a\sin\theta}
\newcommand{\RprS}{R+r\sin\theta}
\newcommand{\ORpaS}{R+a\sin\hat{\theta}}
\newcommand{\half}{\frac{1}{2}}
\newcommand{\quart}{\frac{1}{4}}
\newcommand{\HS}{\hspace{10.0mm}}
\newcommand{\Hs}{\hspace{5.0mm}}
\newcommand{\VS}{\vspace{10.0mm}}
\newcommand{\Vs}{\vspace{5.0mm}}
\begin{document}
\renewcommand{\thefootnote}{\fnsymbol{footnote}}
\title{
	How to construct a coordinate representation
	of a Hamiltonian operator on a torus
}

\author{
	SUMIO ISHIKAWA\footnotemark[1],
	TADASHI MIYAZAKI, \\
	KAZUYOSHI YAMAMOTO\footnotemark[2]
	and
	MOTOWO YAMANOBE\footnotemark[3] \\
	{\it Department of Physics, Science University of Tokyo,} \\
	{\it Kagurazaka, Shinjuku-ku, Tokyo 162, Japan}
}
\footnotetext[1]{E-mail: liszt@grad.ap.kagu.sut.ac.jp}
\footnotetext[2]{E-mail: ymt@grad.ap.kagu.sut.ac.jp}
\footnotetext[3]{E-mail: yamanobe@grad.ap.kagu.sut.ac.jp}

\date{October 1995}

\maketitle
\vspace{20mm}
\begin{flushleft}
\end{flushleft}
\vspace{20mm}

\begin{abstract}
The dynamical system of a point particle constrained on a torus
is quantized {\it \`a la} Dirac
with two kinds of coordinate systems respectively;
the Cartesian and toric coordinate systems.
In the Cartesian coordinate system,
it is difficult to express momentum operators
in coordinate representation
owing to the complication
in structure of the commutation relations
between canonical variables.
In the toric coordinate system,
the commutation relations have a simple form
and their solutions in coordinate representation
are easily obtained with,
furthermore, two quantum Hamiltonians turning up.
A problem comes out when the coordinate system is transformed,
{\it after} quantization,
from the Cartesian to the toric coordinate system.
\end{abstract}

\newpage
\setlength{\baselineskip}{8mm}

\section{Introduction}

How can one formulate, in the Hamiltonian formalism,
the classical system of a particle
constrained on a curved surface?
Starting with a classical-mechanical system
for that constrained particle,
how can one obtain a quantum-mechanical one?

The first question is solved
in use of the Dirac formalism to the system\cite{dirac}.
In the Hamiltonian formalism,
Poisson brackets play a main role
when we deal with a system without constraints.
When we are confronted, on the other hand,
with a constrained system,
the Dirac brackets take their place.
Making use of the Dirac brackets,
we can consistently formulate the constrained system
in an elegant way.

Noticing the point that
the Dirac brackets for the constrained system
play the role of the Poisson brackets for the unconstrained system,
we can quantize the classical constrained system,
i.e., we can answer the second question.
The Dirac brackets are to be replaced, in quantization,
by the commutators ($\times 1/i\hbar$).
Along this canonical-quantization method,
therefore, it is important that we describe the system
in the language of the Hamiltonian formalism,
not of the Lagrangian formalism.

Now,
let us focus our attention on the system of a particle
constrained on a {\it two-dimensional} closed surface.
The simplest closed surface is a sphere.
Quantum mechanics for a particle on a sphere
has been studied by many researchers\cite{ohnuki},
among whom are Falck and Hirshfeld\cite{falck}.

{}From the mathematical viewpoint,
any compact orientable surface is known to be homeomorphic to
a sphere or a connected sum of tori\cite{massey}.
However, the quantum mechanics on a {\it torus} has not,
within the knowledge of the present authors,
been constructed, so we will work that out in the present paper.
We are thus to have quantum mechanics
on all compact orientable surfaces.

The present paper is organized as follows.
In Sec.2
we study the classical system of a particle
on a torus {\it \`a la} Dirac.
Two coordinate systems are used,
one is Cartesian and the other toric.
The reason why the toric coordinate system is useful is
that one can here easily obtain coordinate representations
of momentum operators.
In Sec.3
we quantize the system with the canonical-quantization method.
In the classical theory of a particle constrained on a surface,
the Dirac brackets of the canonical variables
are complicated in structure
if one deals with the system in Cartesian coordinates,
as shown in Sec.2.1.
Therefore, it is difficult
to obtain the coordinate representations of momentum operators.
There are two ways available to avoid that difficulty.
One is
to choose a suitable coordinate system for the dynamical system
in which the Dirac brackets come to have simple structures.
The other was shown by
Homma, Inamoto and Miyazaki in case of a sphere\cite{homma}.
They took up the time derivative of the equation of a surface
as a constraint condition.
We go along the former way in the present paper.
In Sec.3.1 we study quantization in toric coordinates.
Two Hamiltonian operators are shown to exist;
one consists only of differential terms, and the other
both of differential and functional ones.
In Sec.3.2
we first quantize the constrained system in Cartesian coordinates,
and transform the coordinates
to the toric afterwards.
It is found that the momentum operator becomes hermitian or not,
depending on the stage at which
we impose the constraint condition on the system.
The last section is devoted
to the conclusion and discussion of our analysis.

\vspace{10mm}
\setcounter{equation}{0}
\setcounter{section}{1}
\section{Classical mechanics on a torus}

In ${\bf R}^3$ $(x_1,x_2,x_3)$
we consider a particle on a torus
which is generated by rotating a circle
(radius:$a$, center:$(R,0,0)$, in $x_1\mbox-x_3$ plane)
about the $x_3$-axis.
It is expressed by
\begin{equation}
	x_3^2-a^2+(\RXX-R)^2=0.				\label{eqn:0}
\end{equation}
%

\subsection{Classical mechanics in Cartesian coordinates}

The Lagrangian $L$ of this system is given by
\begin{equation}
	L
	=
	\half m\dot{x}_i\dot{x}_i
	- x_4\left(x_3^2-a^2+(\RXX-R)^2\right).		\label{eqn:1}
\end{equation}
where $x_4(t)$ is a Lagrange multiplier and treated as an
independent variable,
and $m$ is the mass of the particle.
Here and henceforth the summation convention
of the repeated indices is employed.
{}From Eq.(\ref{eqn:1}), we obtain canonical momenta:
\begin{equation}
	P_i
	\equiv
	\Diff{L}{\dot{x}_i}
	=
	m\dot{x}_i,
\Hs
        (i=1,2,3),					\label{eqn:2}
\end{equation}
\begin{equation}
	P_4 \equiv \Diff{L}{\dot{x}_4}
	\equiv
	\phi_1 \approx 0,				\label{eqn:3}
\end{equation}
where ``$\approx$'' means weak equality in Dirac's sense.
Equation (\ref{eqn:3}) is the primary constraint of this system.
{}From Eqs.(\ref{eqn:1}) $\sim$ (\ref{eqn:3}), the Hamiltonian $H$ is
given by
\begin{equation}
	H
	=
	\frac{1}{2m}P_iP_i+x_4
	\left[x_3^2 - a^2 + (\RXX-R)^2\right]
	+ uP_4,						\label{eqn:4}
\end{equation}
where $u (\equiv \dot{x_4})$ is a Lagrange multiplier.
To keep consistency for the system,
all constraints are to be imposed
after working out all Poisson brackets.
In order that the system be compatible with the dynamical evolution,
we require all constraints to be conserved throughout all time.
This requirement is called a consistency condition.
In our case,
it is necessary that $\phi_2\equiv\dot{\phi}_1\approx0$.
We have thus a new constraint (secondary constraint)
$\phi_2\approx 0$ on the system.
Furthermore,
we impose the consistency condition on $\phi_2 \approx 0$.
In general, above arguments continue till
either no new constraint turns up further
or a condition on a Lagrange multiplier in the Hamiltonian
is obtained.
In this paper,
the consistency condition finally gives a condition on $u$
in Eq.(\ref{eqn:4}).
\begin{eqnarray}
	\dot{\phi}_1
	& \approx &
	x_3^2-a^2+(\RXX-R)^2
	\equiv
	\phi_2 \approx 0,				\label{eqn:5}
\\
	\dot{\phi}_2
	& \approx &
	x_iP_i-\frac{R}{\RXX}(\XP)
	\equiv
	\phi_3 \approx 0,				\label{eqn:6}
\\
	\dot{\phi}_3
	& \approx &
	\frac{1}{m\RXX}
	\left[
	P_iP_i
	+ \frac{R\left(\XP\right)^2}{\XX}
	- R \left(P_1^2+P_2^2\right)
	\right]					\nonumber
\\
	& &
	+ 2x_4
	\left[
	x_3^2
	+ \left(\RXX-R\right)^2
	\right]
	\equiv
	\frac{\phi_4}{m} \approx 0,			\label{eqn:7}
\\
	\dot{\phi}_4
	& \approx &
	- \frac{3}{mR}\frac{x_iP_i}{\RXX}
	\left(2x_4a^2-\frac{P_jP_j}{m}\right)
	- 2ua^2 \approx 0.				\label{eqn:8}
\end{eqnarray}
Equation (\ref{eqn:8}) determines the Lagrange multiplier $u$
in the Hamiltonian (\ref{eqn:4}).
Note that all constraints are found to be second-class; i.e.
there is no constraint
which has zero Poisson brackets with all other constraints.
Therefore, Eq.(\ref{eqn:4}) is rewritten as
\begin{equation}
	H
	=
	\frac{1}{2m}P_iP_i
	+ x_4\phi_2
	- \frac{3}{2mRa^2}\frac{x_iP_i}{\RXX}
	\left(2x_4a^2-\frac{P_jP_j}{m}\right)\phi_1.	\label{eqn:9}
\end{equation}
At this stage the Poisson bracket is defined by
\begin{equation}
	\PB{A,B}
	\equiv
	\Diff{A}{x_\mu}\Diff{B}{P_\mu}
	- \Diff{A}{P_\mu}\Diff{B}{x_\mu},
\HS
	({\rm summation\  over}\ \mu=1,2,3,4).		\label{eqn:10}
\end{equation}

Equations (\ref{eqn:9}) and (\ref{eqn:10}) have a variable $x_4$
in themselves.
We try to express $x_4$ in terms of other variables.
{}From $\phi_2\approx 0$ and $\phi_4\approx 0$, we obtain
\begin{equation}
	x_4
	\approx
	\frac{1}{2ma^2}
	\left[
	P_iP_i
	+ \frac{R}{\RXX}
	\left(\frac{(\XP)^2}{\XX} - P_1^2 - P_2^2\right)
	\right].					\label{eqn:11}
\end{equation}
With this equation as well as Eq.(\ref{eqn:3}),
we regard the pair $(x_4,P_4)$
as dependent variables, and hence, the remaining constraints
are $\phi_2 \approx 0$ and $\phi_3 \approx 0$.

Now, we introduce the Dirac bracket defined by
\begin{equation}
	\DB{F,G}
	\equiv
	\PB{F,G}
	- \PB{F,\phi_\alpha}
	C^{-1}_{\alpha\beta}
	\PB{\phi_\beta,G},				\label{eqn:12}
\end{equation}
with
\begin{displaymath}
	C_{\alpha \beta}
	\equiv
	\PB{\phi_\alpha ,\phi_\beta},
\Hs
	C_{\alpha \beta}C^{-1}_{\beta \gamma}
	=
	\CD{\alpha \gamma},
\Hs
        (\alpha,\beta,\gamma=1,2).
\end{displaymath}
{}From Eq.(\ref{eqn:12}),
the Dirac brackets of canonical variables are given as follows:
\begin{eqnarray}
	\DB{x_i,x_j}
	& = &
	0,						\label{eqn:14}
\\
	\DB{x_i,P_j}
	& = &
	\CD{ij}
	- \frac{1}{a^2}
	\left(x_i-\frac{\XD{i}}{\RXX}R\right)
	\left(x_j-\frac{\XD{j}}{\RXX}R\right),		\label{eqn:15}
\\
	\DB{P_i,P_j}
	& = &
	-\frac{1}{a^2}
	\left[
	\left(x_i-\frac{\XD{i}}{\RXX}R\right)
	\right.						\nonumber
\\
	& &
	\times
	\left(
	P_j-\frac{R}{\RXX}
	\left(
	P_1\CD{1j}+P_2\CD{2j}-\frac{\XP}{\XX}
	\left(\XD{j}\right)
	\right)
	\right)						\nonumber
\\
	& &
	-\left(
	x_j-\frac{\XD{j}}{\RXX}R
	\right)						\nonumber
\\
	& &
	\left.
	\times
	\left(
	P_i-\frac{R}{\RXX}
	\left(
	P_1\CD{1i}+P_2\CD{2i}-\frac{\XP}{\XX}
	\left(\XD{i}\right)
	\right)
	\right)
	\right].\Hs					\label{eqn:16}
\end{eqnarray}

Making use of the Dirac brackets,
let us replace the weak equality ``$\approx$''
by the strong equality ``$=$''.
Finally the Hamiltonian is reduced to
\begin{equation}
	H_E
	=
	\frac{1}{2m}P_iP_i,
\Hs
        (i=1,2,3).					\label{eqn:17}
\end{equation}
Hamilton's equations of motion are
\begin{eqnarray}
	\dot{x}_i
	& = &
	\DB{x_i,H_E}
	=
	\frac{P_i}{m},					\label{eqn:18}
\\
	\dot{P}_i
	& = &
	\DB{P_i,H_E}
	=
	- 2\alpha
	\left(
	x_i-\frac{\XD{i}}{\RXX}R
	\right),					\label{eqn:19}
\end{eqnarray}
where
\begin{equation}
	\alpha
	\equiv
	\frac{1}{2ma^2}
	\left[
	P_iP_i
	+ \frac{R}{\RXX}
	\left(\frac{(\XP)^2}{\XX}-P_1^2-P_2^2\right)
	\right].					\label{eqn:20}
\end{equation}
In particular,
as $R\rightarrow 0$ in Eqs.(\ref{eqn:19}) and (\ref{eqn:20}),
we obtain
\begin{equation}
	\dot{P}_i
	=
	- \frac{1}{ma^2}P_jP_jx_i
	\equiv
	-\, m\omega^2x_i,
\HS
	\omega:{\rm const.}				\label{eqn:21}
\end{equation}
Equation (\ref{eqn:21}) expresses an equation of circular motion,
as was expected,
with angular frequency $\omega=\sqrt{P_jP_j}/ma$.

Now we can quantize the system
by replacing Dirac brackets ($\times i\hbar$) by commutators.
However,
since the Dirac brackets of the canonical variables are complicated
in structure as shown in Eqs.(\ref{eqn:15}) and (\ref{eqn:16}),
it is very difficult
to obtain the coordinate representation of momentum operators.
We will make no further mention of this difficulty in this section.
In the next subsection,
we define a toric coordinate system
and consider the system in that coordinates.

\subsection{Classical mechanics in toric coordinates}

We introduce a new coordinate system and call it a {\it toric} one.
The variables are defined as illustrated in Fig.1.
The relationship between Cartesian coordinates and toric
is given by
\begin{eqnarray}
	x_1
	& = &
	(\RprS)\cos\phi,				\nonumber
\\
	x_2
	& = &
	(\RprS)\sin\phi,				\label{eqn:22}
\\
	x_3
	& = &
	r\cos\theta,					\nonumber
\end{eqnarray}
with
\begin{displaymath}
	r\sin\theta > -R.
\end{displaymath}

The Lagrangian $L$ in the toric coordinate system is
\begin{equation}
	L
	=
	\half m
	\left[
	\dot{r}^2+r^2\dot{\theta}^2+(\RprS)^2\dot{\phi}^2
	\right]
	- q_4(r-a),					\label{eqn:23}
\end{equation}
where $q_4$ is a Lagrange multiplier.
Canonical momenta are given by
\begin{eqnarray}
	\Pi_r
	& \equiv &
	\frac{\partial L}{\partial \dot{r}}
	=
	m\dot{r},					\label{eqn:24}
\\
	\Pi_\theta
	& \equiv &
	\Diff{L}{\dot{\theta}}
	=
	mr^2\dot{\theta},				\label{eqn:25}
\\
	\Pi_\phi
	& \equiv &
	\Diff{L}{\dot{\phi}}
	=
	m(\RprS)\dot{\phi},				\label{eqn:26}
\\
	\Pi_4
	& \equiv &
	\Diff{L}{\dot{q_4}}
	\equiv
	\chi_1 \approx 0.				\label{eqn:27}
\end{eqnarray}
Equation (\ref{eqn:27}) is the primary constraint of this system.
By the Legendre transformation,
the Hamiltonian $H$ is obtained:
\begin{equation}
	H
	=
	\frac{1}{2m}
	\left[
	\Pi_r^2
	+ \frac{\Pi_\theta^2}{r^2}
	+ \frac{\Pi_\phi^2}{(\RprS)^2}
	\right]
	+ q_4(r-a) + u\chi_1,				\label{eqn:28}
\end{equation}
where $u (\equiv \dot{q_4})$ is a Lagrange multiplier.
At this stage the Poisson bracket is defined by
\begin{equation}
	\PB{A,B}
	\equiv
	\Diff{A}{q_\mu}\Diff{B}{\Pi_\mu}
	- \Diff{A}{\Pi_\mu}\Diff{B}{q_\mu},
\Hs
        (\mu=1,2,3,4).					\label{eqn:29}
\end{equation}
In Eq.(\ref{eqn:29}), we use the following notation:
\begin{displaymath}
	q_1 = r,
\Hs
	q_2 = \theta,
\Hs
	q_3 = \phi,
\end{displaymath}
\begin{displaymath}
	\Pi_1 = \Pi_r,
\Hs
	\Pi_2 = \Pi_\theta,
\Hs
	\Pi_3 = \Pi_\phi,
\end{displaymath}
Requiring the consistency condition on all constraints,
we obtain three secondary constraints,
\begin{eqnarray}
	\dot{\chi}_1
	& \approx &
	r-a \equiv \chi_2 \approx 0,			\label{eqn:30}
\\
	\dot{\chi}_2
	& \approx &
	\frac{\Pi_r}{m}
	\equiv
	\frac{\chi_3}{m} \approx 0,			\label{eqn:31}
\\
	\dot{\chi}_3
	& \approx &
	\frac{1}{m}
	\left[
	\frac{\Pi_\theta^2}{a^3}
	+ \frac{\Pi_\phi^2\sin\theta}{(\RpaS)^3}-mq_4
	\right]
	\equiv \frac{\chi_4}{m} \approx 0,		\label{eqn:32}
\\
	\dot{\chi}_4
	& \approx &
	\frac{\Pi_\theta\Pi_\phi^2\cos\theta}{a(\RpaS)^3}
	\left(\frac{1}{a}-\frac{\sin\theta}{\RpaS}\right)
	- \frac{m^2}{3}u \approx 0.			\label{eqn:33}
\end{eqnarray}
Equation (\ref{eqn:33}) determines the Lagrange multiplier $u$;
therefore Eq.(\ref{eqn:28}) is rewritten as
\begin{eqnarray}
	H
	& = &
	\frac{1}{2m}
	\left[
	\Pi_r^2+\frac{\Pi_\theta^2}{r^2}
	+ \frac{\Pi_\phi^2}{(\RprS)^2}
	\right]
	+q_4(r-a)					\nonumber
\\
	& &
	+ \frac{3\Pi_\theta\Pi_\phi^2\cos\theta}{am^2(\RpaS)^3}
	\left(\frac{1}{a}-\frac{\sin\theta}{\RpaS}\right)
	\chi_1.						\label{eqn:34}
\end{eqnarray}

We express $q_4$ in term of other variables by Eq.(\ref{eqn:32}):
\begin{equation}
	q_4
	\approx
	\frac{1}{m}
	\left[
	\frac{\Pi_\theta^2}{a^3}
	+\frac{\Pi_\phi^2\sin\theta}{(\RpaS)^3}
	\right].					\label{eqn:35}
\end{equation}
According to the same argument as in Sec.2.1,
the pair $(q_4,\Pi_4)$ are taken as dependent variables.
Substituting Eq.(\ref{eqn:35}) into Eq.(\ref{eqn:34}),
the Hamiltonian (\ref{eqn:34}) is reduced to
\begin{eqnarray}
	H
	& = &
	\frac{1}{2m}
	\left[
	\Pi_r^2 + \frac{\Pi_\theta^2}{r^2}
	+ \frac{\Pi_\phi^2}{(\RprS)^2}
	\right]
	+ \frac{1}{m}
	\left[
	\frac{\Pi_\theta^2}{a^3}
	+ \frac{\Pi_\phi^2\sin\theta}{(\RpaS)^3}
	\right]
	(r-a)						\nonumber
\\
	& &
	+ \frac{3\Pi_\theta\Pi_\phi^2\cos\theta}{m^2r(\RprS)^3}
	\left(\frac{1}{r}-\frac{\sin\theta}{\RprS}\right)
	\chi_1.						\label{eqn:36}
\end{eqnarray}
and here the Poisson bracket is redefined by
\begin{equation}
	\PB{A,B}
	\equiv
	\Diff{A}{q_i}\Diff{B}{\Pi_i}
	- \Diff{A}{\Pi_i}\Diff{B}{q_i},
\Hs
        (i=1,2,3).					\label{eqn:37}
\end{equation}
Now,
we calculate the Dirac brackets
of canonical variables $(q_i,\Pi_i,i=1,2,3)$:
\begin{equation}
	\DB{\theta,\Pi_\theta}
	=
	\DB{\phi,\Pi_\phi}
	= 1,						\label{eqn:39}
\end{equation}
and all other Dirac brackets vanish.
{}From now on,
since the system is taken up through the Dirac brackets,
we can replace ``$\approx$'' by ``$=$''.
Finally we arrive at
\begin{equation}
	H_E
	=
	\frac{1}{2m}
	\left[
	\frac{\Pi_\theta^2}{a^2}
	+ \frac{\Pi_\phi^2}{(\RpaS)^2}
	\right],					\label{eqn:40}
\end{equation}
with
\begin{displaymath}
	r = a,
	\HS
	\Pi_r = 0.
\end{displaymath}
Hamilton's equations of motion are given as follows:
\begin{eqnarray}
	\dot{r} = 0,
\hspace{9.35em}
	& &
	\dot{\Pi}_r = 0,				\label{eqn:41}
\\
	\dot{\theta} = \frac{\Pi_\theta}{ma^2},
\hspace{7.85em}
	& &
	\dot{\Pi}_\theta
	=
	- \frac{a\cos\theta\,\Pi_\phi^2}{m(\RpaS)^3},	\label{eqn:42}
\\
	\dot{\phi} = \frac{\Pi_\phi}{m(\RpaS)^2},
\hspace{3.0em}
	& &
	\dot{\Pi}_\phi = 0.				\label{eqn:43}
\end{eqnarray}

In the above discussion,
we treat $(r,\theta,\phi)$ as coordinate variables.
However we can reduce them to $(\theta,\phi)$
by solving the equations
$\chi_2=0$ and $\chi_3=0$ for $r$ and $\Pi_r$.
In this case, since all constraints are disappeared,
we can deal with the system through the Poisson brackets
constructed by  $(\theta,\phi,\Pi_\theta,\Pi_\phi)$.
Nevertheless,
the resultant Hamiltonian is the same as Eq.(\ref{eqn:40}).
\vspace{10mm}

\setcounter{equation}{0}
\section{Quantum mechanics on a torus}

Now, we consider the quantization of
the constrained system discussed in the previous section.
In Sec.3.1,
we quantize the system in toric coordinates.
In Sec.3.2,
we bring toric coordinates in
after having quantized in Cartesian coordinates.

\subsection{Quantization in toric coordinates}

Following Eq.(\ref{eqn:40}),
we define the quantum Hamiltonian $\hat{H}$ as:
\begin{equation}
	\hat{H}
	=
	\frac{1}{2m}
	\left[
	\frac{\hat{\Pi}_\theta^2}{a^2}
	+ \frac{\hat{\Pi}_\phi^2}{(\ORpaS)^2}
	\right],					\label{eqn:44}
\end{equation}
with
\begin{displaymath}
	\hat{r} = a,
\HS
	\hat{\Pi}_r = 0,
\end{displaymath}
where we put the notation ``$\ \hat{}\ $''
to stress the operator nature of the affixed.
The commutation relations of the canonical operators is
\begin{equation}
	[\,\hat{q}_m,\hat{\Pi}_n] = i\hbar\CD{mn},
\HS
	(m,n = 2,3).					\label{eqn:45}
\end{equation}
All other commutators vanish.

According to
the representation theory in a general coordinate system
developed by De Witt\cite{dewitt},
we will rewrite Eq.(\ref{eqn:44})
in coordinate representation.
We start with a brief review.

On an n-dimensional Riemannian manifold,
we suppose to have the commutators of
the coordinate operators
and their canonical-momentum operators
$(\hat{x}_\mu,\hat{P}_\mu),\ \mu=1,2,...,n$
as follows:
\begin{equation}
	[\,\hat{x}^\mu,\hat{x}^\nu]
	=
	[\hat{P}_\mu,\hat{P}_\nu]
	= 0,
\HS
	[\,\hat{x}^\mu,\hat{P}_\nu]
	=
	i\hbar\CD{\nu}^\mu.				\label{eqn:46}
\end{equation}
The wave function of the system is defined by:
\begin{equation}
	\Psi(x,t) \equiv \langle x|\Psi\rangle_t,       \label{eqn:47}
\end{equation}
where $|x\rangle$ is an eigenvector of the operator $\hat{x}$
and satisfies the orthonormal condition
\begin{equation}
	\langle x |\, x' \rangle = \delta(x,x').	\label{eqn:48}
\end{equation}
The generalized delta function $\delta(x,x')$
means the equations:
\begin{equation}
	\delta(x,x')=0,
\HS
	(x \neq x'),
           						\label{eqn:49}
\end{equation}
\begin{equation}
	\int d\omega f(x)\delta(x,x')=f(x'),
\HS
	d\omega \equiv g^{\half}(x)d^nx,		\label{eqn:50}
\end{equation}
with $g(x)$,
the determinant of the metric.
It is related to Dirac's delta function by:
\begin{equation}
	\delta(x,x')
	= g^{\half}(x)\,\delta(x-x')
	= g^{\half}(x')\,\delta(x-x').			\label{eqn:51}
\end{equation}
{}From Eq.(\ref{eqn:51}) follow two identities:
\begin{eqnarray}
	(x^\mu-x'^\mu)\Diff{}{x^\nu}\delta(x,x')
	& = &
	-\,\delta_\nu^\mu\delta(x,x'),			\label{eqn:52}
\\
	\Diff{}{x^\mu}\delta(x,x')
	& = &
	-\Diff{}{x'^\mu}\delta(x,x')
	-\Gamma^\nu_{\nu\mu}\delta(x,x'),		\label{eqn:53}
\end{eqnarray}
where $\Gamma^\mu_{\nu\lambda}$ is the Cristoffel symbol calculated
at $x$ or $x'$.
With the commutators (\ref{eqn:46}),
we obtain the equations:
\begin{equation}
	i\hbar\,\delta^\mu_\nu\delta(x,x')
	=
	(x^\mu-x'^\mu)
	\langle x|\hat{P}_\nu|\,x'\rangle,		\label{eqn:54}
\end{equation}
and
\begin{equation}
	\langle x|\hat{P}_\nu|\,x'\rangle
	=
	- i\hbar\Diff{}{x^\nu}\delta(x,x')
	+ F_\nu\delta(x,x').				\label{eqn:55}
\end{equation}
Here $F_\nu$ is an arbitrary function of $x$ and $x'$.

On account that the momentum operators $\hat{P}_\mu$
commute with each other
and that $\hat{P}_\mu$ should be Hermitian operator,
we obtain
\begin{eqnarray}
	F_\mu
	& = &
	 \Diff{}{x^\mu}F,				\label{eqn:56}
\\
	F
	& \equiv &
	 R-\half\,i\hbar\log g^{\half},			\nonumber
\end{eqnarray}
with an arbitrary real function $R$.
The function $R$ can be eliminated by unitary transformation:
\begin{equation}
	|\,x\rangle '
	=
	\exp\left[-\frac{i}{\hbar}R\right]|\,x\rangle.	\label{eqn:57}
\end{equation}
Then we have the coordinate representation of $\hat{P}_\mu$:
\begin{equation}
	\langle x|\hat{P}_\mu|\,x'\rangle
	=
	-i\hbar\Diff{}{x^\mu}\delta(x,x')
	-\half i\hbar\Gamma^\nu_{\nu\mu}\delta(x,x').	\label{eqn:58}
\end{equation}

Now having finished a brief overview,
we will rewrite the momentum operators
$\hat{\Pi}_\theta$, $\hat{\Pi}_\phi$,
using Eq.(\ref{eqn:58}),
in coordinate representation.
The surface element on torus is
\begin{displaymath}
	ds^2
	=
	a^2d\theta^2+(\RpaS)^2d\phi^2
	\equiv
	g_{mn}dq_mdq_n,
\HS
	(m,n=2,3).
\end{displaymath}
It shows that the metric on torus is
\begin{eqnarray}
	g_{mn}
	& \equiv &
	\left(
	\begin{array}{cc} a^2 & 0 \\ 0 & (\RpaS)^2 \end{array}
	\right),
\\
	g_{mn}g^{-1}_{nl}
	& = &
	\CD{ml}.					\label{eqn:59}
\end{eqnarray}
We obtain the commutation relations of the canonical variables
by replacing the Dirac brackets ($\times i\hbar$)
by commutators in Eq.(\ref{eqn:39}):
\begin{equation}
	[\,\hat{q}_m,\hat{\Pi}_n]
	=
	i\hbar\,\CD{mn},
\HS
	(m,n = 2,3),					\label{eqn:60}
\end{equation}
and all other commutators vanish.
Then the momentum operators are expressed as:
\begin{eqnarray}
	\hat{\Pi}_\theta
	& = &
	\frac{\hbar}{i}
	\left(
	\Diff{}{\theta}
	+ \half\frac{a\cos\theta}{\RpaS}
	\right),        				\label{eqn:61}
\\
	\hat{\Pi}_\phi
	& = &
	\frac{\hbar}{i}\Diff{}{\phi}.			\label{eqn:62}
\end{eqnarray}

Now, by substituting Eqs.(\ref{eqn:61}), (\ref{eqn:62})
into Eq.(\ref{eqn:44}),
we obtain the quantum Hamiltonian $\hat{H}$:
\begin{eqnarray}
	\hat{H}
	& = &
	\frac{1}{2m}
	\left[
	\frac{\hat{\Pi}_\theta^2}{a^2}
	+ \frac{\hat{\Pi}_\phi^2}{(\ORpaS)^2}
	\right]						\nonumber
\\
	& = &
	-\frac{\hbar^2}{2m}
	\left[
	\frac{1}{a^2(\RpaS)}\Diff{}{\theta}
	\left(
        (\RpaS)\Diff{}{\theta}
	\right)
	+ \frac{1}{(\RpaS)^2}\Diff{{}^2}{\phi^2}
	\right]						\nonumber
\\
	& &
	+ \frac{\hbar^2}{8m}
	\frac{2a^2+2aR\sin\theta-a^2\cos^2\theta}
	{a^2(\RpaS)^2}.					\label{eqn:63}
\end{eqnarray}
In the last line of Eq.(\ref{eqn:63}),
a functional term has appeared.
It is called the quantum mechanical potential (QMP),
having been indicated by De Witt.
On the contrary,
we alternatively find that we can use the Hamiltonian
\begin{equation}
	\hat{H}
	=
	\frac{1}{2m}
	\hat{g}^{-\quart}\hat{\Pi}_m\,
	\hat{g}^{\half}\hat{g}_{mn}^{-1}
	\,\hat{\Pi}_n\,\hat{g}^{-\quart},		\label{eqn:64}
\end{equation}
\begin{displaymath}
	\hat{g}
	\equiv
	\det |\hat{g}_{mn}|,
\HS
	(m,n=2,3),
\end{displaymath}
instead of the Hamiltonian (\ref{eqn:44}).
Then the coordinate representation of
the Hamiltonian (\ref{eqn:64}) is obtained as follows:
\begin{equation}
	\hat{H}
	=
	- \frac{\hbar^2}{2m}
	\left[
	\frac{1}{a^2(\RpaS)}\Diff{}{\theta}
	\left((\RpaS)\Diff{}{\theta}\right)
	+ \frac{1}{(\RpaS)^2}\Diff{{}^2}{\phi^2}
	\right].					\label{eqn:65}
\end{equation}

We have no reason to decide
which Hamiltonian (\ref{eqn:44}) or (\ref{eqn:64}) be preferable.
This problem of selection of preferable Hamiltonian
is well known to appear
if one wants to use the polar-coordinate representation
in three-dimensional space.
There is, however, no theoretical rule for the selection.
\Vs

\subsection{Quantization in Cartesian coordinates}

In this subsection,
we first quantize in the Cartesian coordinate system.
We then rewrite in the toric coordinate system,
point-transforming the former.
The Dirac brackets (\ref{eqn:14}) and (\ref{eqn:15})
are replaced by commutators ($\times 1/i\hbar$)
\begin{eqnarray}
	[\,\hat{x}_i,\hat{x}_j]
	& = &
	0,						\label{eqn:66}
\\
	\left[\,\hat{x}_i,\hat{P}_j\right]
	& = &
	i\hbar\,\CD{ij}
	-\frac{i\hbar}{a^2}
	\left(
	\hat{x}_i
	- \frac{\hat{x}_1\CD{1i}+\hat{x}_2\CD{2i}}{\ORXX}R
	\right)
	\left(
	\hat{x}_j
	- \frac{\hat{x}_1\CD{1j} +\hat{x}_2\CD{2j}}{\ORXX}R
	\right).					\label{eqn:67}
\end{eqnarray}
In these equations,
it is confirmed that
$\hat{x}_i$ and $\hat{P}_i$ are Hermitian operators.
Because of the Hermiticity of $\hat{P}_i$,
the commutators between them
obtained from Eq.(\ref{eqn:16}) should be
\begin{eqnarray}
	\left[\hat{P}_i,\hat{P}_j\right]
	& = &
	- \frac{i\hbar}{a^2}
	\left[
	\left(\hat{x}_i-\frac{\OXD{i}}{\ORXX}R\right)
	\right.						\nonumber
\\
	& &
	\left.\times
	\left(
	\hat{P}_j-\frac{R}{\ORXX}
	\left(
	\hat{P}_1\CD{1j}+\hat{P}_2\CD{2j}
	- \frac{\OXD{j}}{\OXX}\left(\OXP\right)
	\right)
	\right)
	\right.						\nonumber
\\
	& &
	\left.
	- \left(\hat{x}_j-\frac{\OXD{j}}{\ORXX}R\right)
	\right.						\nonumber
\\
	& &
	\left.\times
	\left(
	\hat{P}_i-\frac{R}{\ORXX}
	\left(
	\hat{P}_1\CD{1i}+\hat{P}_2\CD{2i}
	- \frac{\OXD{i}}{\OXX}\left(\OXP\right)
	\right)
	\right)
	\right].\Hs					\label{eqn:68}
\end{eqnarray}
Furthermore,
we symmetrize the constraint conditions:
\begin{equation}
	\hat{x}_3^2-a^2
	+ \left(\ORXX-R\right)^2
	=
	0,						\label{eqn:69}
\end{equation}
\begin{equation}
	\half
	\left(
	\hat{x}_i\hat{P}_i+\hat{P}_i\hat{x}_i
	\right)
	- \half R
	\left[
	\frac{1}{\ORXX}
	\left(\hat{x}_1\hat{P}_1 + \hat{x}_2\hat{P}_2\right)
	+
	\left(\hat{P}_1\hat{x}_1+\hat{P}_2\hat{x}_2\right)
	\frac{1}{\ORXX}
	\right]
	=
	0.						\label{eqn:70}
\end{equation}

Then we define the point transformation
of the Cartesian coordinate system to the toric one.
The toric coordinate variables defined by Eq.(\ref{eqn:22}) are
expressed in the Cartesian coordinates as:
\begin{eqnarray}
	q_1
	& \equiv &
	r = x_3^2+(\RXX-R)^2,				\nonumber
\\
	q_2
	& \equiv &
	\theta = \tan^{-1}\frac{\RXX-R}{x_3},		\label{eqn:71}
\\
	q_3
	& \equiv &
	\phi = \tan^{-1}\frac{x_2}{x_1},		\nonumber
\end{eqnarray}
with
\begin{displaymath}
	r\sin\theta > -R,
\end{displaymath}
for one-to-one correspondence.
At the classical level,
the momentum variables in both coordinate systems
are related by the point transformation as follows:
\begin{eqnarray}
	\Pi_i
	& = &
	\Diff{x_j}{q_i}P_j,		 		\label{eqn:72}
\\
	P_i
	& = &
	\Diff{q_j}{x_i}\Pi_j.				\label{eqn:73}
\end{eqnarray}
The point transformation is available at the quantum level
by symmetrizing:
\begin{eqnarray}
	\hat{\Pi}_i
	& = &
	\half
	\left[
	\Diff{\hat{x}_j}{\hat{q}_i}\hat{P}_j
	+ \hat{P}_j \Diff{\hat{x}_j}{\hat{q}_i}
	\right],					\label{eqn:74}
\\
	\hat{P_i}
	& = &
	\half
	\left[
	\Diff{\hat{q}_j}{\hat{x}_i}\hat{\Pi}_j
	+ \hat{\Pi}_j \Diff{\hat{q}_j}{\hat{x}_i}
	\right].					\label{eqn:75}
\end{eqnarray}
Substituting Eqs. (\ref{eqn:71}) and (\ref{eqn:75})
into Eqs. (\ref{eqn:66})--(\ref{eqn:68}),
we obtain the commutators of the canonical operators
in toric coordinates:
\begin{equation}
	[\,\hat{q}_m,\hat{\Pi}_n]=i\hbar\,\CD{mn},
\HS
	(m,n=2,3),					\label{eqn:76}
\end{equation}
and all other commutators vanish.
This equation is just the same as Eq.(\ref{eqn:60}),
which means that the quantization at the different stage,
i.e.,
at the direct stage of the toric coordinate representation
or at the stage of the Cartesian coordinate representation
(afterwards point-transforming
to the toric coordinate representation),
has no influence on the resultant commutators.
The constraints (\ref{eqn:69}) and (\ref{eqn:70})
are rewritten as:
\begin{equation}
	\hat{r}^2=a^2,
\HS
	\half
	\left(\hat{\Pi}_r\hat{r}+\hat{r}\hat{\Pi}_r\right)
	=
	0.						\label{eqn:77}
\end{equation}

Next,
we will rewrite the quantum Hamiltonian $\hat{H}$
in toric coordinates.
The Hamiltonian quantized in Cartesian coordinates
is given by
\begin{equation}
	\hat{H}
	=
	\frac{1}{2m}\hat{P}_i\hat{P}_i.			\label{eqn:78}
\end{equation}
The coordinate transformation in the Hamiltonian
is made by
substitution of Eqs. (\ref{eqn:71}) and (\ref{eqn:75})

The momentum operators $\hat{P}_i$ are expressed as:
\begin{eqnarray}
	\hat{P}_i
	& = &
	\left(
	\hat{\Pi}_j
	\left(\Diff{\hat{q}_j}{\hat{x}_i}\right)
	+ \half
	\left[
	\left(\Diff{\hat{q}_j}{\hat{x}_i}\right),\hat{\Pi}_j
	\right]
	\right)						\nonumber
\\
	& = &
	\left(
	\hat{\Pi}_j
	+ \frac{i\hbar}{2}
	\left(\Diff{\hat{x}_k }{\hat{q}_l}\right)
	\left(\Diff{}{\hat{q}_j}\Diff{\hat{q}_l}{\hat{x}_k}\right)
	\right)
	\left(\Diff{\hat{q}_j}{\hat{x}_i}\right)	\nonumber
\\
	& = &
	\hat{g}^{-\quart}\hat{\Pi}_j\hat{g}^{\quart}
	\left(\frac{\hat{q}_j}{\hat{x}_i}\right),
\HS
	(i,j,k=1,2,3).					\label{eqn:79}
\end{eqnarray}
In the process
from the second line to the third of Eq.(\ref{eqn:79}),
we have made use of the matrix identity
\begin{equation}
	{\rm tr}[M^{-1}\partial_kM]
	=
	\left(\det M\right)^{-1}
	\partial_k \det M.				\label{eqn:80}
\end{equation}

Now we introduce new momentum operators $\hat{P'}_i$,
which are defined as the Cartesian momentum operators
restricted to the constraint condition (\ref{eqn:77}).
They are expressed as:
\begin{equation}
	\hat{P'}_i
	=
	\sum_{m=2}^3\hat{g}^{-\quart}\hat{\Pi}_m\hat{g}^{\quart}
	\left(\Diff{\hat{q}_m}{\hat{x}_i}\right),	\label{eqn:81}
\end{equation}
and interpreted as the projected Cartesian momentum operators
onto the torus,
and should be observable.

We will check the Hermiticity of $\hat{P'}_3$.
By Eq.(\ref{eqn:81}), $\hat{P'}_3$ is
\begin{eqnarray}
	\hat{P'}_3
	& = &
	- \frac{1}{a}\sin\theta\hat{\Pi}_\theta
	- \frac{3\hbar}{2ia}\cos\theta			\nonumber
\\
	& = &
	- \frac{\hbar}{i}
	\left[
	\frac{1}{a}\sin\theta\Diff{}{\theta}
	+ \frac{1}{a}\cos\theta
	+ \frac{\sin\theta\cos\theta}{\RpaS}
	\right].					\label{eqn:82}
\end{eqnarray}
If we substitute $\Pi_r=0$ into Eq.(\ref{eqn:75}),
we have
\begin{eqnarray}
	\hat{P''}_3
	& = &
	\half\sum_{m=2}^3
	\left[\Diff{\hat{q}_m}{\hat{x}_3}\hat{\Pi}_m
	+ \hat{\Pi}_m\Diff{\hat{q}_m}{\hat{x}_3}
	\right]						\nonumber
\\
	& = &
	-\frac{\hbar}{i}
	\left[
	\frac{1}{a}\sin\theta\Diff{}{\theta}
	+ \frac{1}{2a}\cos\theta
	+ \half\frac{\sin\theta\cos\theta}{\RpaS}
	\right].					\label{eqn:83}
\end{eqnarray}
The result does not coincide with the above equation (\ref{eqn:82}).
In Eqs.(\ref{eqn:82}) and (\ref{eqn:83}),
we find that
${P''}_3$ is a Hermitian operator, while ${P'}_3$ is not.
This teaches us
to put the restriction $\hat{\Pi}_r=0$ on the Hamiltonian
{\it after} the substitution of
Eq.(\ref{eqn:79}) into Eq.(\ref{eqn:78}).
We arrive at the resultant Hamiltonian
\begin{eqnarray}
	\hat{H}
	& = &
	\frac{1}{2m}\hat{P}_i\hat{P}^{\dag}_i		\nonumber
\\
	& = &
	\frac{1}{2m}
	\hat{g}^{-\quart}\hat{\Pi}_m\,
	\hat{g}^{\half}\hat{g}_{mn}^{-1}\,
	\hat{\Pi}_n\,\hat{g}^{-\quart}.			\label{eqn:84}
\end{eqnarray}
The Hamiltonian (\ref{eqn:84}) has no QMP
in contrast to Eq.(\ref{eqn:64}),
and is a  Hermitian operator.
We finally obtain the coordinates representation of the Hamiltonian,
whose system has first been quantized in Cartesian coordinates
and point-transformed into toric coordinates afterward.
Namely we thus have
\begin{equation}
	\hat{H}
	=
	- \frac{\hbar^2}{2m}
	\left[
	\frac{1}{a^2(\RpaS)}\Diff{}{\theta}
	\left((\RpaS)\Diff{}{\theta}\right)
	+ \frac{1}{(\RpaS)^2}\Diff{{}^2}{\phi^2}
	\right].					\label{eqn:85}
\end{equation}
%

\setcounter{equation}{0}
\section{Conclusion}

We have quantized our constrained system based on
the Dirac formalism for the classical constrained system
as well as on the canonical-quantization method.
A preferable quantum Hamiltonian for
a particle constrained on a torus has been obtained.

In quantum theory,
to express the momentum operators
in coordinate representation
and to have the Schr\"{o}dinger equation,
the commutators of canonical variables play an essential role.
The commutators
(\ref{eqn:66})--(\ref{eqn:68})
in the Cartesian coordinate system
are much complicated.
Therefore,
it is very difficult
to obtain the coordinate representation of
the momentum operators in this system.
Using the toric coordinate system enables us
to have simple commutators
and represent the momentum operators
as operators in coordinate space.
Although the toric coordinate system defined by Eq.(\ref{eqn:22})
has a restriction $r\sin\theta > -R$,
it is adequate
to deal with the constrained system of a particle on the torus.
With this coordinate system we have two Hamiltonians
Eqs.(\ref{eqn:63}) and (\ref{eqn:65}),
the former with QMP and the latter without QMP.
With which Hamiltonian
(\ref{eqn:44}) or (\ref{eqn:64}) we start,
in momentum representation, gives rise to this difference.

In quantum theory,
the order that we first make a coordinate transformation
then restrict the system under some constraint conditions,
or vice versa,
gives a great influence on the resultant theory.
We conclude, in this paper, that,
under the condition $\hat{\Pi}_r=0$,
the point transformations (\ref{eqn:74}) and (\ref{eqn:75})
are quite misleading.
We must first transform the Hamiltonian
into the form shown in Eq.(\ref{eqn:84}),
then afterwards restrict it
to the constraint condition $\hat{\Pi}_r=0$.
We thus have a preferable
quantum Hamiltonian Eq.(\ref{eqn:85}).
\VS

\noindent
{\Large\bf Acknowledgement} \\
Two of the authors (K.Y. and M.Y.) would like to thank Iwanami F\^{u}jukai
for financial support.


\end{document}